\documentclass[aps,pra,reprint,amsmath,amssymb,superscriptaddress,showpacs,nobibnotes]{revtex4-1}
\usepackage{graphicx,SIunits}
\usepackage{bm}     
\usepackage{hyperref} 
\usepackage{dsfont}    
\usepackage{color}
\DeclareMathOperator{\Tr}{Tr}

\begin{document}


\title{Robustness of reference-frame-independent quantum key distribution \\
against the relative motion of the reference frames}



\author{Tanumoy Pramanik}
\email{tanu.pra99@gmail.com}
\affiliation{Center for Quantum Information, Korea Institute of Science and Technology (KIST), Seoul, 02792, Republic of Korea}

\author{Byung Kwon Park}
\affiliation{Center for Quantum Information, Korea Institute of Science and Technology (KIST), Seoul, 02792, Republic of Korea}
\affiliation{Department of Nano-Materials Science and Engineering, Korea University of Science and Technology, Daejeon, 34113, Republic of Korea}

\author{Young-Wook Cho}
\affiliation{Center for Quantum Information, Korea Institute of Science and Technology (KIST), Seoul, 02792, Republic of Korea}

\author{Sang-Wook Han}
\affiliation{Center for Quantum Information, Korea Institute of Science and Technology (KIST), Seoul, 02792, Republic of Korea}

\author{Sang-Yun Lee}
\affiliation{Center for Quantum Information, Korea Institute of Science and Technology (KIST), Seoul, 02792, Republic of Korea}

\author{Yong-Su Kim}
\email{yong-su.kim@kist.re.kr}
\affiliation{Center for Quantum Information, Korea Institute of Science and Technology (KIST), Seoul, 02792, Republic of Korea}
\affiliation{Department of Nano-Materials Science and Engineering, Korea University of Science and Technology, Daejeon, 34113, Republic of Korea}

\author{Sung Moon}
\affiliation{Center for Quantum Information, Korea Institute of Science and Technology (KIST), Seoul, 02792, Republic of Korea}

\date{\today} 

\begin{abstract}
\noindent Reference-Frame-Independent quantum key distribution (RFI-QKD) is known to be robust against slowly varying reference frames. However, other QKD protocols such as BB84 can also provide secrete keys if the speed of the relative motion of the reference frames is slow enough. While there has been a few studies to quantify the speed of the relative motion of the reference frames in RFI-QKD, it is not yet clear if RFI-QKD provides better performance than other QKD protocols under this condition. Here, we analyze and compare the security of RFI-QKD and BB84 protocol in the presence of the relative motion of the reference frames. In order to compare their security in real world implementation, we also consider the QKD protocols with decoy state method. Our analysis shows that RFI-QKD provides more robustness than BB84 protocol against the relative motion of the reference frames.
\end{abstract}


\maketitle
\section{Introduction}

\noindent Quantum key distribution (QKD) promises enhanced communication security based on the laws of quantum physics~\cite{BB84, Ekert91}. Since the first QKD protocol has been introduced in 1984, there has been a lot of theoretical and experimental effort to improve the security and the practicality of QKD~\cite{gisin02, scarani09}. These days, QKD research is not only limited in laboratories~\cite{lo12,patel14,guan15,jeong16,choi16} but also in industries~\cite{QKD_commer}.

In general, QKD requires a shared common reference frame between two communicating parties, Alice and Bob. For example, the interferometric stability or the alignment of the polarization axes are required for fiber based QKD using phase encoding and polarization encoding free-space QKD, respectively. However, it can be difficult and costly to maintain the shared reference frame in real world implementation. For instance, it is highly impractical to establish a common polarization axes in earth-to-satellite QKD due to the revolution and rotation of the satellite with respect to the ground station~\cite{rarity02, ursin07,schmitt07,bonato09,meyer11,bourgoin13,bourgoin15}. 

A recently proposed reference-frame-independent QKD (RFI-QKD) provides an efficient way to bypass this shared reference frame problem~\cite{laing10}. In RFI-QKD, Alice and Bob share the secrete keys via a decoherence-free basis while check the communication security with other bases. Both free-space~\cite{wabnig13} and telecom fiber~\cite{zhang14,liang14} based RFI-QKD have been successfully implemented. It is remarkable that the concept of the reference frame independent can be applied to measurement-device-independent QKD~\cite{yin14, wang15}. 

Unlike to its name, however, the security of the original theory of RFI-QKD is guaranteed when the relative motion of the reference frames is slow comparing to the system repetition rate~\cite{laing10}. If the reference frames of Alice and Bob are deviated with a fixed angle, however, one can easily compensate the deviation and implement an ordinary QKD protocol. Therefore, the effectiveness of RFI-QKD over other QKD protocols becomes clear when there is rapid relative motion of the reference frames during the QKD communication. There has been few studies to quantify the speed of the relative motion of the reference frames in RFI-QKD~\cite{sheridan10,wang16}. Without the performance comparison with other QKD protocols, however, these studies do not show the effectiveness of RFI-QKD over other QKD protocols.

In this paper, we report the security of RFI-QKD and BB84 protocol in the presence of the relative motion of the reference frames of Alice and Bob. In order to compare the performances in real world implementation, we also consider the decoy state method. By comparing the security analyses, we found that RFI-QKD is more robust than BB84 protocol against the relative motion of the reference frames.


\section{QKD with a fixed \\reference frame deviation}

In this section, we review the security proof of RFI-QKD and BB84 protocol with a fixed reference frame deviation. A shared reference frame is required for both fiber based QKD with phase and free-space QKD with polarization encoding. It corresponds to the interferometric stability and the polarization axes for fiber based QKD and free-space QKD, respectively. In the following, we will consider free-space QKD with polarization encoding for simplicity. However, we note that our analysis is also applicable for fiber based QKD with phase encoding.

\subsection{RFI-QKD protocol}



\begin{figure}[b]
\includegraphics[width=2.5in]{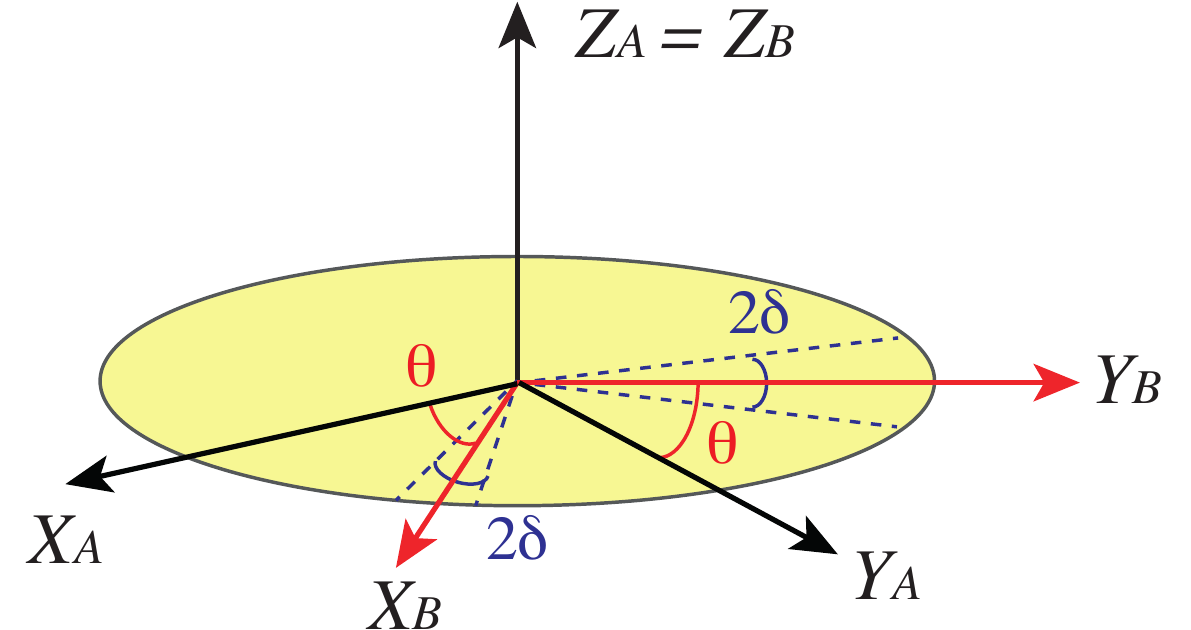}
\caption{The polarization axes of Alice and Bob. $\theta$ and $2\delta$ denote the fixed angle deviation, and the range of the relative motion of the polarizations axes, respectively.} 
\label{Fig_1}
\end{figure}

Figure~\ref{Fig_1} shows the polarization axes of Alice and Bob with a deviation angle $\theta$. Since Alice and Bob should face each other in order to transmit the optical pulses, their $Z$-axes, which corresponds to left- and right-circular polarization states, are always well aligned. On the other hand, the relation of their $X$ and $Y$-axes, which correspond to linear polarization states such as horizontal, vertical, $+45\degree$, and $-45\degree$ polarization states, depend on $\theta$. The relations of the polarization axes are
\begin{eqnarray}
X_B&=&X_A\cos\theta+Y_A\sin\theta, \nonumber\\
Y_B&=&Y_A\cos\theta-X_A\sin\theta, \nonumber\\
Z_B&=&Z_A,
\label{rotation}
\end{eqnarray}
where the subscripts $A$ and $B$ denote Alice and Bob, respectively. 

In RFI-QKD, Alice and Bob share the secrete keys via $Z$-axis, as it is unaffected by the polarization axes deviation. In this basis, the quantum bit error rate (QBER) becomes
\begin{equation}
Q_{ZZ}=\frac{1-\langle Z_AZ_B \rangle}{2}.
\label{QBER_RFI}
\end{equation}
Here, the subscripts $ij$ where $i,j\in\{X,Y,Z\}$ denote that Alice sends a state in $i$ basis while Bob measures it in $j$ basis. The probability distributions of the measurement outcomes in $X$ and $Y$-axes are used to estimate the knowledge of an eavesdropper, Eve. Her knowledge can be estimated by a quantity $C$ which is defined as
\begin{equation}
C= \langle X_A\,X_B\rangle^2 + \langle X_A\,Y_B\rangle^2 + \langle Y_A\,X_B\rangle^2 + \langle Y_A\,Y_B\rangle^2.
\label{C_RFI}
\end{equation}
Note that the quantity $C$ is independent of the deviation angle $\theta$.
The knowledge of Eve is bounded by
\begin{equation}
I_E\left[Q_{ZZ},C\right] = (1-Q_{ZZ}) H\left[ \frac{1+u}{2}\right] + Q_{ZZ} H\left[\frac{1+v}{2}\right],
\end{equation}
where 
\begin{eqnarray}
u &=& \min\left[ \frac{1}{1-Q_{ZZ}} \sqrt{\frac{C}{2}},1 \right], \nonumber \\
v&=&\frac{1}{Q_{ZZ}} \sqrt{\frac{C}{2} - (1-Q_{ZZ})^2 u^2},
\end{eqnarray}
and  $H\left[x\right]=-x\log_2x - (1-x)\log_2(1-x)$ is the Shannon entropy of $x$. 

The secret key rate in the RFI-QKD protocol is given by~\cite{laing10}
\begin{eqnarray}
r_{RFI} = 1-H\left[Q_{ZZ}\right] - I_E\left[Q_{ZZ},C\right].
\label{r_RFI}
\end{eqnarray}
It is notable that Eq.~(\ref{r_RFI}) is independent of a fixed deviation rotation $\theta$~\cite{laing10}. The security proof shows that $r_{RFI}\geq0$ for $Q_{ZZ}\lesssim15.9\%$.

In practice, the effective quantum state that Bob receives from Alice can have errors due to the transmission noise and experimental imperfection. Assuming the noise and the imperfection are polarization independent, we can model the Bob's receiving quantum state $\rho_B$ as 
\begin{eqnarray}
\rho_B= p\rho_A + \frac{1-p}{2}I,
\label{rho_B}
\end{eqnarray}
where $\rho_A$, $1-p$ and $I$ are the state prepared by Alice, the strength of noise, and a two dimensional identity matrix, respectively. Note that $\langle\mathcal{F}_A\mathcal{G}_B\rangle$ can be written as a state dependent form of
$\langle\mathcal{F}_A\mathcal{G}_B\rangle=\Tr[(\mathcal{F}_A\otimes\mathcal{G}_B)\cdot\rho_{AB}]$
where $\mathcal{F,G}\in\{X,Y,Z\}$, and $\rho_{AB}=\rho_A\otimes\rho_B$. Therefore, the QBER $Q_{ZZ}$ and the quantity $C$ become
\begin{eqnarray}
Q_{ZZ}&=&\frac{1-p}{2},\label{C_RFI_rho_B}
\label{Q_Z_p}\\
C&=&2\,p^2= 2 (1-2 Q_{ZZ})^2.
\end{eqnarray}
In this case, $r_{RFI}\geq0$ for $Q_{ZZ} \lesssim 12.6 \%$. 

\subsection{BB84 protocol}

In this section, we consider the secrete key rate of BB84 with a fixed reference frame deviation. Due to the symmetry, the QBER of $X$, and $Y$-axes are the same, and they are given as
\begin{eqnarray}
Q_{XX}&=&\frac{1-\langle X_AX_B\rangle}{2},\nonumber\\
&=&\frac{1-p\cos\theta}{2}=Q_{YY}.
\label{BB84_QBER_X}
\end{eqnarray}
If Alice and Bob utilize $X$ and $Y$ axes, the overall QBER $Q_{\overline{XY}}$ is 
\begin{eqnarray}
Q_{\overline{XY}} &=& \frac{1}{2}\left(Q_{XX}+Q_{YY}\right) \nonumber \\
&=& \frac{1}{2} \left( 1- p \cos\theta\right).
\label{Q_XY_Av}
\end{eqnarray}
Since we know that $Z$-axis is rotation invariant, one can get lower QBER by using $Z$-axis instead of $Y$-axis. In this case, the overall QBER $Q_{\overline{XZ}}$ is given by
\begin{eqnarray}
Q_{\overline{XZ}} &=& \frac{1}{2}\left(Q_{XX}+Q_{ZZ}\right) \nonumber \\
&=& \frac{1}{2} \left( 1- p\cos^2\frac{\theta}{2}\right).
\label{Q_XZ_Av}
\end{eqnarray}

The secrete key rate of BB84 with $\{X,Y\} (\{X,Z\})$ bases is given by~\cite{gisin02,scarani09} 
\begin{equation}
r^{XZ\,(XY)}_{BB84} = 1- 2 H\left[Q_{\overline{XZ}\,(\overline{XY})}\right].
\label{r_BB84}
\end{equation}
Apparently, Eq.~(\ref{r_BB84}) is dependent on the reference frame deviation $\theta$. However, one can easily compensate the deviation if $\theta$ is invariant during the QKD communication. For BB84 protocol, $r^{XZ\,(XY)}_{BB84}\geq0$ for $Q_{ZZ} \lesssim 11\%$ when $\theta=0$.

\section{QKD in the presence of the relative motion of reference frames}

In this section, we study the effect of the relative motion of the reference frames of Alice and Bob during the QKD communication. Let us consider the case when $\theta$ varies from $-\delta$ to $\delta$ as depicted in Fig.~\ref{Fig_1}. For simplicity, we assume that $\theta$ is centered at 0 and equally distributed over $\theta\in\left[-\delta,\delta\right]$.

The quantities $Q_{XX}$, $Q_{YY}$, and $C$ are affected by the relative motion of the reference frames. However, $Q_{ZZ}$ is unchanged since $Z_A=Z_B$ all the time. In order to quantify the effect of the relative motion, we need to calculate the average values of the observed quantities of ${\langle \mathcal{F}_A\mathcal{G}_B\rangle}$, which is given by
\begin{eqnarray}
\overline{\langle \mathcal{F}_A\mathcal{G}_B\rangle} &=& \frac{1}{2\delta} \int_{-\delta}^{\delta} \, \langle \mathcal{F}_A\mathcal{G}_B\rangle d\theta \nonumber \\
&=&\frac{\sin2\delta}{\delta} \langle \mathcal{F}_A\mathcal{G}_A\rangle.
\label{frame_deviation}
\end{eqnarray}

With Eq.~(\ref{frame_deviation}), one can represent the quantity $C$ as a function of $Q_{ZZ}$ and $\delta$, 
\begin{equation}
C\left[\,Q_{ZZ},\delta\,\right] = 2 (1-2 Q_{ZZ})^2 \left(\frac{\sin2\delta}{2 \delta}\right)^2.
\label{C_Q/delta}
\end{equation}
Therefore, one can estimate the secrete key rate of RFI-QKD in the presence of the relative motion of the reference frames by inserting Eq.~(\ref{C_RFI_rho_B}) and (\ref{C_Q/delta}) to Eq.~(\ref{r_RFI}).

\begin{figure}[t]
\includegraphics[width=3.5in]{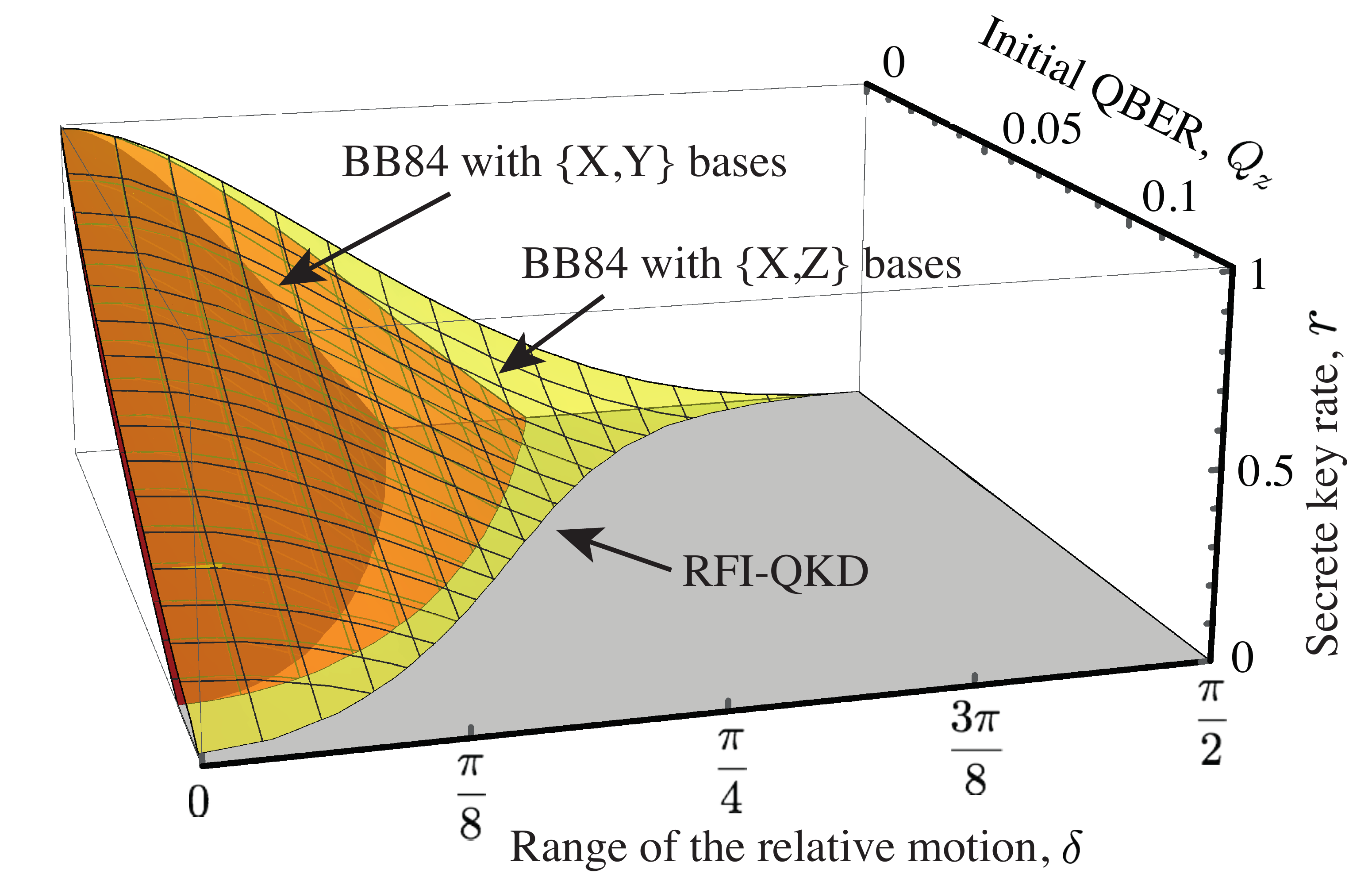}
\caption{The lower bounds of the secrete key rates of RFI-QKD and BB84 protocol with $\{X,Y\}(\{X,Z\})$ bases with respect to $Q_{ZZ}$ and $\delta$.}
\label{Fig_2}
\end{figure}

The average QBER $Q_{XX}$ and $Q_{YY}$ are also presented as a function of $Q_{ZZ}$ and $\delta$, and they become
\begin{eqnarray}
Q_{XX}\left[\,Q_{ZZ},\delta\,\right]
&=&Q_{YY}\left[\,Q_{ZZ},\delta\,\right]\nonumber\\
&=& \frac{1}{2\delta} \int_{-\delta}^{\delta} \frac{1-\langle X_AX_B\rangle}{2} d\theta \nonumber \\
&=& \frac{1}{2} - \frac{(1-2 Q_{ZZ}) \sin2\delta}{4 \delta}.
\end{eqnarray}
Therefore, the average QBER $Q_{\overline{XY}}$ and $Q_{\overline{XZ}}$ become
\begin{eqnarray}
Q_{\overline{XY}}\left[\,Q_{ZZ},\delta\,\right] &=&\frac{1}{2}\left(Q_{XX}\left[\,Q_{ZZ},\delta\,\right] +  Q_{YY}\left[\,Q_{ZZ},\delta\,\right]\right), \nonumber \\
&=& \frac{1}{2} -  \frac{(1-2Q_{ZZ}) \sin2\delta}{4\delta}.
\label{Q_XY}
\end{eqnarray}
and
\begin{eqnarray}
Q_{\overline{XZ}}\left[\,Q_{ZZ},\delta\,\right]  &=& \frac{1}{2}\left(Q_{XX}\left[\,Q_{ZZ},\delta\,\right] +  Q_{ZZ}\left[\,Q_{ZZ},\delta\,\right]\right), \nonumber \\
&=& \frac{1+2Q_{ZZ}}{4} - \frac{(1-2Q_{ZZ}) \sin2\delta}{8\delta}.
\label{Q_XZ}
\end{eqnarray}
By inserting Eq.~(\ref{Q_XY}) or (\ref{Q_XZ}) to Eq.~(\ref{r_BB84}), one can estimate the secrete key rate of BB84 either with $\{X,Y\}$ or $\{X,Z\}$ bases in the presence of the relative motion of the reference frames.

In Fig.~\ref{Fig_2}, we compare the lower bounds of the secrete key rates for RFI-QKD and BB84 with $\{X,Y\}$ and $\{X,Z\}$ bases with respect to $Q_{ZZ}$ and $\delta$. It clearly shows that RFI-QKD is more robust than BB84 protocol against the channel depolarization as well as the relative motion of the reference frames. In the case of BB84 protocol, one can obtain better results by using $\{X,Z\}$ bases instead of $\{X,Y\}$ bases.


\section{RFI-QKD and BB84 protocol using decoy state method}

In this section, we apply the security analysis to a real world implementation using weak coherent pulses with decoy state method. In the following, we will consider two decoy states method which is most widely used for real world implementation~\cite{Decoy2}. According to this method, Alice randomly modulates the intensity of the weak coherent pulses with mean photon numbers per pulse of $\mu$, $\nu$ and and 0 where $\mu>\nu$. They are usually called signal, decoy and vacuum pulses, respectively. 

Assuming that the bases chosen by Alice and Bob are $i$ and $j$ where $i,j\in\{X,Y,Z\}$, the lower bound of the secrete key rate is given by~\cite{Decoy,Decoy2}
\begin{equation}
r = -Y_{\mu} \; H[Q_{ij|\mu}] + \mu \; y_1^L \exp[-\mu]\left( 1- I_E\right).
\label{r_Decoy}
\end{equation}
where $Y_{\mu}$, and $y_1^L$ are the gain of signal pulses having QBER $Q_{ij|\mu}$, and the lower bound of the gain of single-photon pulses, repectively. Here, we assume that the error correction efficiency is unity. The values of $Y_{\mu}$ and $Q_{ij|\mu}$ can be obtained from the experiment, however, $y_1^L$ should be estimated with the decoy and vacuum pulses. The lower bound of the gain of single-photon pulses is given by~\cite{liang14}
\begin{eqnarray}
y_1^L = \frac{\mu^2 Y_{\nu} e^{\nu}-\nu^2 Y_{\mu} e^{\mu} - \left( \mu^2-\nu^2\right)Y_0}{\mu\left(\mu \nu -\nu^2\right)},
\label{S_Ph_C_Rt}
\end{eqnarray}
where $Y_{\nu}$  and $Y_0$ are the gain of decoy pulses and vacuum pulses, respectively. 

The QBER for signal state $Q_{ij|\mu}$ can be calculated by using
\begin{equation}
Y_{\mu}\,Q_{ij|\mu} = \sum_{n=0}^{\infty} y_n \, \frac{\mu^n\, q_{n,ij}}{n!} \exp[-\mu],
\label{QmuEmu}
\end{equation}
 where $y_n$, and $q_{n,ij}$ are the gains and QBER of $n$-photon state, respectively. Here, $Y_{\mu}$ and $q_{n,ij}$ are given as
\begin{eqnarray}
Y_{\mu} &=& \sum_{i=0}^{\infty} y_i\, \frac{\mu^i}{i!}\,\exp\left[-\mu\right], \nonumber \\
&=& 1-\exp\left[-\eta \,\mu\right] + p_d\label{Y_mu}
\\
q_{n,ij}&=&\frac{e_{ij} \, \eta_n + \frac{1}{2} p_d}{y_n},
\label{QBER_nP}
\end{eqnarray}
where $\eta$, $p_d$, and $e_{ij}$ denote the overall detection efficiency including the channel transmission, the dark count probability, and the erroneous detection probability, respectively. The detection efficiency of $n$-photon state $\eta_n$ is given by $\eta_n=1-(1-\eta)^n$ where $\eta$ can be represented in terms of loss, $L$ in dB and Bob's detection efficiency, $\eta_B$ by $\eta=\eta_B10^{-\frac{L}{10}}$.

The deviation between the reference frames of Alice and Bob contributes the erroneous detections, hence, $e_{ij}=Q_{ij}$. Note that  $Q_{XY}=\frac{1}{2\delta} \int_{-\delta}^{\delta}\frac{1-\langle X_A Y_B\rangle}{2}d\theta= \frac{1}{2}$, $Q_{YX}=\frac{1}{2\delta} \int_{-\delta}^{\delta} \frac{1-\langle Y_A X_B\rangle}{2}d\theta=\frac{1}{2}$, and $Q_{ii}$ are given by Eqs.~(\ref{Q_Z_p}), and (\ref{BB84_QBER_X}).

According to the decoy theory, the upper bound of QBER generated by single-photon states of the signal pulses is given by~\cite{liang14}
\begin{eqnarray}
q_{1,ij|\mu}^{U} = \frac{Q_{\mu ij} Y_{\mu} - \frac{1}{2}Y_0\exp[-\mu]}{\mu y_1^L \exp[-\mu]}.
\label{UB_QBER_1}
\end{eqnarray}
The lower bound of QBER occurred due to single-photon states of the signal pulses is given by 
\begin{eqnarray}
q_{1,ij|\mu}^L= 1-\frac{\left(1-Q_{\mu\, ij}\right)Y_\mu-\frac{1}{2} Y_0 \exp[-\mu]}{\mu y_1^L\exp[\mu]}.
\label{LB_QBER_1}
\end{eqnarray}
Here, we assume that the QBER of $n$-photon states for $n\geq 2$ is $q_{n,ij}=1$.

\subsection{RFI-QKD protocol using decoy state}

In the RFI-QKD scenario, the lower bound of the secrete key rate $r_{RFI}$ becomes~\cite{liang14}
\begin{equation}
r_{RFI} = -Y_{\mu} \; H[Q_{ZZ|\mu}] + \mu \; y_1^L \exp[-\mu]\left( 1- I_E\right).
\label{r_RFI_De}
\end{equation}
The upper bound of Eve's information $I_E$ is given by
\begin{eqnarray}
I_E&=& \left(1-q_{1,ZZ|\mu}^U\right)\,H\left[\frac{1+u_{\max}}{2}\right] \nonumber \\
&&- q_{1,ZZ|\mu}^U\,H\left[\frac{1+v}{2}\right].
\end{eqnarray}
where $q_{1,ZZ|\mu}^U$ given by Eq.~(\ref{UB_QBER_1}), and 
\begin{eqnarray}
u_{\max} &=& \min\left[\frac{1}{1-q_{1,ZZ|\mu}^U} \sqrt{\frac{C_1^L}{2}},\,1\right], \nonumber \\
v &=& \frac{1}{q_{1,ZZ|\mu}^U} \sqrt{\frac{C_1^L}{2}-\left(1-q_{1,ZZ|\mu}^U\right)^2\,u_{\max}^2}.
\end{eqnarray}
Here, $C_1^L$ is the optimal lower bound of $C$ for single-photon states. The optimal lower bound $C_1^L$ is given below~\cite{liang14}
\begin{eqnarray}
C_1^L = \max[\alpha,\,2(\alpha^\prime)^2] + \max[\beta,\,2(\beta^\prime)^2],
\end{eqnarray}
where 
\begin{eqnarray}
\alpha &=& \sum_{j=X,Y} \left(1-2\max\left(1/2,q_{1\,\mu\,Xj}^L\right)\right)^2, \nonumber \\
\beta&=&\sum_{j=X,Y}\left(1-2\max\left(1/2,q_{1\,\mu\,Yj}^L\right)\right)^2, \nonumber \\
\alpha^\prime &=& \frac{(1.70711-Q_{\mu\,XX}-Q_{\mu\,XY})Y_{\mu}-0.70711 Y_0 \exp[-\mu] }{\mu y_1^L} \nonumber \\
 && - 0.70711, \nonumber \\
\beta^\prime &=& \frac{(1.70711-Q_{\mu\,YX}-Q_{\mu\,YY})Y_{\mu}-0.70711 Y_0 \exp[-\mu] }{\mu y_1^L} \nonumber \\
&& -0.70711.
\end{eqnarray}

\begin{figure*}[t]
\includegraphics[width=5.5in]{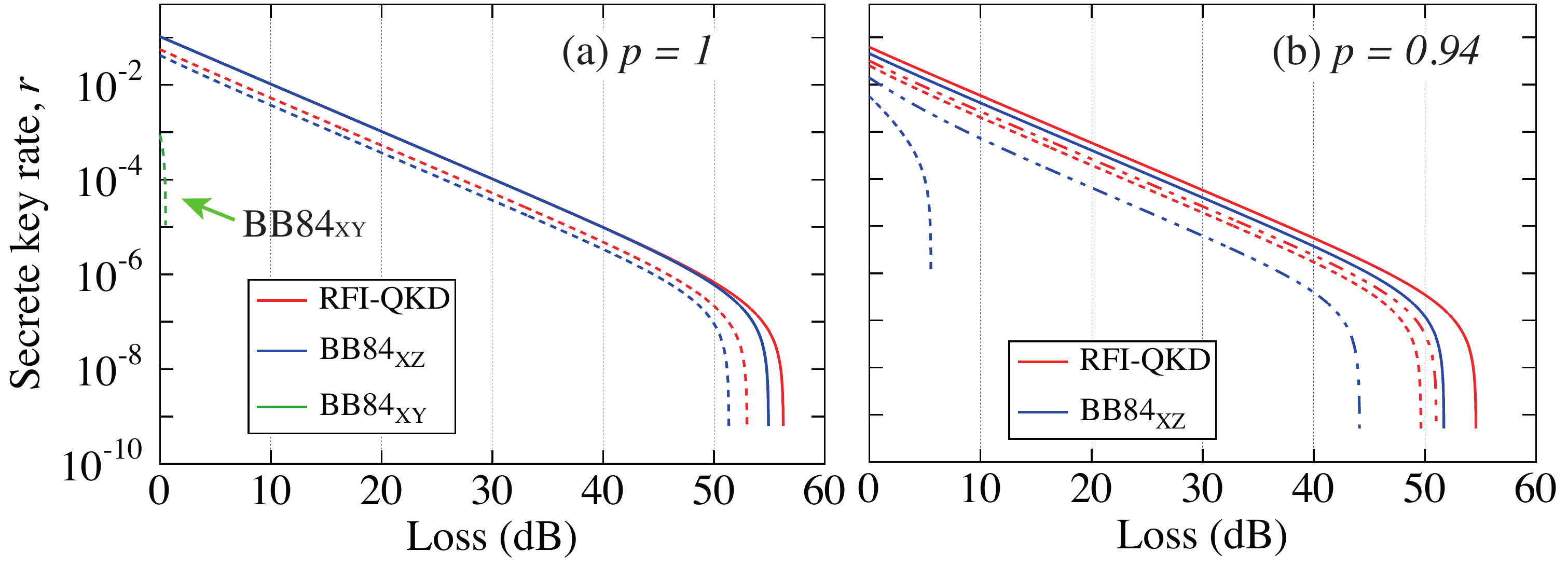}
\caption{The lower bounds of the secrete key rates $r_{RFI}$, $r_{BB84}^{XZ}$, and $r_{BB84}^{XY}$ with respect to loss (a) for ideal case of no noise or $p=1$, and (b) for more realistic case of the intrinsic QBER of $Q_{ZZ}=3\%$ or $p=0.94$. The red, blue and green lines represent $r_{RFI}$, $r_{BB84}^{XZ}$, and $r_{BB84}^{XY}$, respectively. The solid, small-dashed, and dotted lines are for the range of the relative motion of the reference frames $\delta=0, \frac{\pi}{8}$ and $\frac{\pi}{7}$, respectively. For both cases, the secrete key rate of RFI-QKD is more robust than it of BB84 against the relative motion.}
\label{Fig_3}
\end{figure*}

\subsection{BB84 protocol using decoy state}


For BB84 protocol, the lower bounds of the secret key rate with $\{X,\,Z\}$ ($\{X,\,Y\}$) basis is given as~\cite{Decoy,Decoy2}
\begin{eqnarray}
r_{BB84}^{XZ\,(XY)}&=& -Y_{\mu} \; H[Q_{\overline{XZ}|\mu \,(\overline{XY}|\mu) }] 
\label{R_D_XZ_XY} \\
&& + \mu \; y_1^L \exp[-\mu]\left( 1- H\left[q_{1,\overline{XZ}|\mu \,(1,\overline{XY}|\mu)}^U\right]\right).\nonumber
\end{eqnarray}
Note that $q_{1,\overline{XZ}|\mu \,(1,\overline{XY}|\mu)}^U$ is provided at Eq.~(\ref{UB_QBER_1}), and $Q_{\overline{XZ}|\mu}$ and $Q_{\overline{XY}|\mu}$ can be calculated with the help of Eqs.~(\ref{Q_XY_Av}), (\ref{Q_XZ_Av}), (\ref{QmuEmu}), (\ref{Y_mu}), and (\ref{QBER_nP}).


\subsection{Result and discussion}

Figure~\ref{Fig_3} shows the lower bounds of the secret key rates of RFI-QKD and $BB84$ protocols with respect to loss. For simulation, we choose $Y_0=p_d=10^{-6}$, $\mu=0.5$, $\nu=0.05$, and $\eta_B=0.35$ that are widely accepted for the earth-to-satellite QKD scenario~\cite{bourgoin13,bourgoin15}.

Figure~\ref{Fig_3}(a) shows the secrete key rates when there is no intrinsic QBER due to the noise, i.e., $p=1$. The solid red, blue and green lines are for RFI-QKD, BB84 with $\{X,Z\}$ and $\{X,Y\}$ bases with $\delta=0$, respectively. The dotted lines are for $\delta=\frac{\pi}{7}$. When there is no relative motion between the reference frames of Alice and Bob, i.e., $\delta=0$, the secrete ket rates for all QKD protocols are comparable. As the range of the relative motion increases, however, the secrete key rate of BB84 with $\{X,Y\}$ basis rapidly decreases.

Figure~\ref{Fig_3}(b) shows a more realistic case that there is intrinsic QBER due to the noise and the experimental imperfection. Here, we assume the intrinsic QBER of $Q_{ZZ}=3\%$ which corresponds to $p=0.94$. The solid, small-dashed and dotted lines are for $\delta=0, \frac{\pi}{8}$ and $\frac{\pi}{7}$, respectively. It clearly shows that RFI-QKD outperforms BB84 protocol in real world implementation. For example, for $p=0.94, \delta=\frac{\pi}{7}$, the maximum loss that QKD communication is possible in RFI-QKD is reduced by $12\%$ from the ideal case of $p=1,\,\delta=0$. On the other hand, the maximum loss for BB84 protocol using $\{X,\,Z\}$ and $\{X,Y\}$ bases is reduced by $90\%$ and $100\%$ from the ideal case, respectively. 

\section{Conclusions}

To summary, we have studied both reference frame independent quantum key distribution (RFI-QKD) and BB84 protocol with the presence of the relative motion between the reference frames of Alice and Bob. We have also considered overall noise model with a depolarizing channel between Alice and Bob. In order to compare the secrete key rates in real world implementation, we also have applied the security analyses to the decoy state methods. We found that RFI-QKD provides more robustness than BB84 protocol in the presence of the relative motion between the reference frames.

\section*{Acknowledgement}

This work was supported by the ICT R$\&$D program of MSIP/IITP (B0101-16-1355), and the KIST research programs (2E27231, 2V05340).

\end{document}